\documentclass[aps,english,showpacs,10pt]{revtex4}

\usepackage[latin1]{inputenc}
\usepackage{float}
\usepackage{latexsym}
\usepackage{graphicx}
\usepackage{amssymb}
\usepackage{amsmath}
\usepackage{amsfonts}
\usepackage{hyperref}
\usepackage{color}
\usepackage{cool}

\def\be{\begin{equation}} 
\def\ee{\end{equation}} 
\def\bea{\begin{eqnarray}} 
\def\eea{\end{eqnarray}} 

\begin{document}

\title{The Matter Bounce Alternative to Inflationary Cosmology}

\author{Robert Brandenberger}
\email{rhb@physics.mcgill.ca}
\affiliation{Department of Physics, McGill University, Montr\'eal, QC, H3A 2T8, Canada}
\date{\today}

\begin{abstract}

A bouncing cosmology with an initial matter-dominated phase of contraction during
which scales which are currently probed with cosmological observations exit the Hubble radius
provides a mechanism alternative to inflation for producing a nearly scale-invariant
spectrum of cosmological perturbations. In this review article I first discuss the evolution
of cosmological fluctuations in the matter bounce scenario, and then discuss various attempts
at realizing such a scenario. Observational signatures which will allow the matter bounce
to be distinguished from the inflationary paradigm are also discussed.

\end{abstract}

\maketitle

\section{Introduction}

The idea that instead of originating from a Big Bang singularity, the universe has
emerged from a cosmological bounce has a long history (see \cite{Novello}
for a review with an extensive list of references to the original literature). The
group of Professor Novello has made a lot of important contributions to the
research on this topic. However, it was realized only fairly recently \cite{Wands, Fabio1}
that a bouncing cosmology with a matter-dominated phase of contraction during
which scales which are probed today in cosmological observations exit the
Hubble radius can provide an alternative to the current inflationary universe paradigm
of cosmological structure formation. In this review article, we provide an
overview of this ``Matter Bounce" scenario of structure formation, and we discuss
some recent efforts at obtaining a non-singular bouncing cosmological
background (see also \cite{RHBprev} for reviews comparing the Matter Bounce
with inflation and other alternatives to inflation).

Inflationary cosmology \cite{Guth} (see also \cite{Brout, Sato, Starob1}) has become
the paradigm of early universe cosmology not only because it addresses some
of the conceptual problems of Standard Big Bang cosmology such as the
horizon, flatness and entropy problems, but because it provided the first
causal mechanism for generating the primordial fluctuations which could have
developed into the structures we see today on large scales \cite{Mukh} (see also
\cite{Press, Sato, Starob2}). More specifically, it predicted a roughly scale-invariant
spectrum of cosmological fluctuations which in simple models of inflation
have a slight red tilt and which are Gaussian and nearly adiabatic. These predictions
were spectacularly confirmed in recent precision observations of Cosmic Microwave
Background (CMB) anisotropies \cite{WMAP}.

On the other hand, it was known since long before the development of inflationary
cosmology that any model which produces a roughly scale-invariant and almost
adiabatic spectrum of cosmological perturbations will be a good match to observations
\cite{Peebles, Sunyaev}. Inflationary cosmology is simply the first model based on
fundamental physics which yielded such a spectrum. In the mean time, other models
have been developed which predict this kind of spectrum, e.g. the Ekpyrotic
universe \cite{Ekp}, ``string gas cosmology" \cite{SGC}, the ``Varying Speed of Light"
proposal \cite{Moffat}, the ``Conformal Universe" \cite{Rubakov}, and 
the Matter Bounce scenario.

In the following section, I shall show how quantum vacuum fluctuations originating on
sub-Hubble scales and exiting the Hubble radius in a matter-dominated phase of
contracting develop into a scale-invariant spectrum of curvature perturbations. 

As is well known since the pioneering work of Hawking and Penrose (see e.g.
\cite{HE} for a textbook description)), a cosmological singularity is unavoidable if
space-time is described in terms of General Relativity (GR) and if matter obeys
certain energy conditions. Thus, in order to obtain a bouncing cosmology it is
necessary to either go beyond Einstein gravity, or else to introduce new forms
of matter which violate the key energy conditions. In Section 3 of this review
I will describe some concrete and recent models which yield a bouncing
cosmology (the reader is referred to \cite{Novello} for an overview of early
work on obtaining bouncing universes).

\section{Scale-Invariant Fluctuations from a Matter Bounce}

\subsection{Background}

The space-time background cosmology which we have in mind has time $t$
running from $- \infty$ to $\infty$.  The bounce point can be taken to be
$t = 0$. For negative times the universe is contracting. In the absence
of entropy production at the bounce point it is logical to assume that the
contracting phase is the mirror inverse of the expanding cosmology of Standard 
Big Bang cosmology, i.e. at very early times the universe is dominated by
pressure-less matter, and at a time $t = - t_{eq}$ (where $t_{eq}$ is the
time of equal matter and radiation in the expanding phase) there is a
transition to a radiation-dominated phase. If there is entropy production
at the bounce, then the transition from matter to radiation domination will
occur closer to the bounce point than $- t_{eq}$. 

\begin{figure}
\includegraphics[height=.3\textheight]{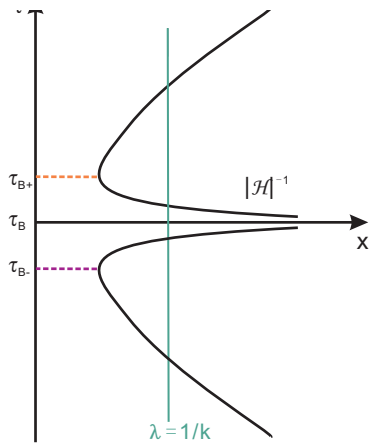}
\caption{Space-time sketch of a non-singular Matter Bounce. The vertical axis
is conformal time, the horizontal axis corresponds to comoving spatial
coordinates. The vertical line indicates the wavelength for some fixed
perturbation mode. The comoving Hubble radius is shown.}
\end{figure}

In Figure 1 we show
a space-time sketch of the bouncing cosmology background. The horizontal
axis is the comoving spatial coordinate, the vertical is time.  The
vertical line indicates the wavelength of a fluctuation model. It should
be compared with the comoving Hubble radius ${\cal H}^{-1}$ (labelled as such
in the figure). As will be reviewed in the following subsection, the Hubble
radius is the maximal scale on which causal and local microphysics
can generate fluctuations.  In order to have a causal generation mechanism,
for fluctuations, it is thus important that wavelengths which are observed today
originate on sub-Hubble scales. The first important point to take away from the
figure is that fixed comoving scales start out with a wavelength smaller than the
Hubble radius and that hence a causal generation mechanism for
fluctuations is possible in a bouncing cosmology, like it is in inflationary
cosmology. In inflationary cosmology it is the exponential decrease
of the comoving Hubble radius during the inflationary phase which allows
for a causal generation mechanism. We see that in a contracting universe
dominated by regular matter a similar decrease in the comoving Hubble
radius occurs.

In this section we discuss how a scale-invariant spectrum of adiabatic
curvature fluctuations emerges from initially vacuum perturbations. 
Whereas in inflationary cosmology there is a symmetry argument
(time translation invariance during the inflationary phase) which
underlies the scale-invariance of the cosmological fluctuations, in a
contracting universe there is no such symmetry. As we will see below,
it is only vacuum fluctuations which exit the Hubble radius in a phase
of matter domination which end up with a scale-invariant spectrum.

\subsection{Preliminaries}

Before studying the generation and evolution of fluctuations
in the Matter Bounce model, it is important to discuss to which
extent this bouncing cosmology can address the conceptual
problems of Standard Big Bang cosmology which the inflationary
scenario successfully addresses, namely the ``Horizon", ``Flatness", ``Size"
and ``Entropy" problems.

In the context of Standard Big Bang cosmology there is no
causal explanation for the observed overall isotropy of
the cosmic microwave background (CMB) since at the time of
last scattering regions in the sky separated by more than
a couple of degrees were causally disconnected \cite{Guth}.
This is the ``Horizon" problem.
Inflation solves this problem by providing a mechanism to
exponentially increase the causal horizon in the early
universe. In a bouncing cosmology the particle horizon
is infinite since time extends to $- \infty$. Hence, the
entire spatial hypersurface at the time of recombination
which we observe today in the CMB is within the causal
horizon and there is no Horizon problem.

In Standard Big Bang cosmology spatial flatness is an
unstable fixed point under dynamical evolution. Thus,
the observed degree of spatial flatness requires extreme
fine tuning of the initial conditions. During an inflationary
phase spatial flatness becomes an attractor, thus
mitigating the Flatness problem. A bouncing cosmology
is ``neutral" with respect to the Flatness problem: if we
postulate a similar degree of spatial flatness an equal
amount of time prior to the bounce point as is observed
today a certain time interval after the bounce point, then
the observations can be explained.

The Size and Entropy problems of the Standard Big Bang
model are based on the idea that the early universe
should be described by initial conditions containing
no hierarchy of scales: at the Planck temperature the
size of space should be Planck size and it should contain
a Planckian value of the entropy. With these initial
conditions inflation is required to generated the
large entropy and size observed today. In a bouncing
cosmology the universe starts large and cool. Thus, there
are no Size and Entropy problems.

There is another aspect of the Flatness problem which
is not addressed in most bouncing cosmologies, namely
the anisotropy problem: having a nearly homogeneous
and isotropic bounce requires fine tuning of the
initial anisotropies. In inflationary cosmology the
anisotropies decrease, and thus the anisotropy problem
is addressed. Similarly, in String Gas Cosmology
there is no anisotropy problem, as shown in \cite{Watson}.
In the Ekpyrotic bouncing cosmology \cite{Ekp} 
anisotropies decrease in importance in the contracting
phase. A way of realizing the Matter Bounce in a
framework which is safe from the anisotropy problem
is to merge ideas from simple Matter Bounce models
with ideas from Ekpyrotic cosmology \cite{Cai3}.
  
\subsection{Formalism}

The fluctuations which we are interested in are inhomogeneities in the
curvature of space-time which are induced by perturbations in matter.
Fluctuations observed today on large cosmological scales are small
in relative amplitude. Hence, we can describe them by linearizing
the Einstein equations about the cosmological background solution.
The corresponding theory is called the ``theory of cosmological
perturbations" (see \cite{MFB} for an in-depth survey
and \cite{RHBrev} for an overview). In the following we summarize
the essentials.

Fluctuations in the space-time metric in cosmology can be classified
according to how they transform under spatial rotations  There are scalar, 
vector and tensor fluctuations. In total there are ten modes. However, four of the
modes correspond to coordinate transformation and are hence
not physical. There are only two scalar, two vector and two tensor
modes which are physical. The tensor modes correspond to
gravitational waves and will not concern us here. Vector modes do
not couple at linear order to matter perturbations and hence will
not be discussed here. In an expanding universe, vector modes decay,
and this yields an additional reason to neglect them. However, in
a contracting universe they grow \cite{Thorsten}, and thus
neglecting them is only justifiable at linear order.The modes of interest
to us are the scalar modes, the ones which at linear order couple
of matter inhomogeneities. For simple forms of matter such as scalar
fields and perfect fluids there is at linear order no anisotropic stress,
and this eliminates one of the scalar modes, leaving one degree of
freedom which describes curvature fluctuations. In longitudinal
gauge, the metric including the scalar fluctuation mode $\Phi(t, x)$
can be written as
\be
ds^2 \, = \, a^2(\eta)\bigl[(1 + 2 \Phi(\eta, {\bf x}))d\eta^2
- (1 - 2 \Phi(t, {\bf x})) d{\bf x}^2 \bigr] \, ,
\ee
where $a(\eta)$ is the cosmological scale factor and we have
made use of conformal time $\eta$ related to physical time $t$
via $a)\eta) d\eta = dt$.

By expanding the full action (Einstein action plus action for matter) to
second order about the classical background cosmology, one can
obtain the action for cosmological perturbations which yields the
linearized equations of motion. This action can be canonically
quantized. By the logic of the previous paragraph, the resulting
action can be written in terms of a single dynamical variable which, in
addition, can be canonically normalized. The form of the canonical fluctuation
variable $v$ was derived by Mukhanov \cite{Mukh2} and Sasaki \cite{Sasaki}
(see also \cite{Lukash}).  The action for $v$ takes the following form
\be \label{pertact}
S^{(2)} \, = \, {1 \over 2} \int d^4x \bigl[v'^2 - v_{,i} v_{,i} + 
{{z''} \over z} v^2 \bigr] \, ,
\ee
where a prime denotes the derivative with respect to conformal time, the
subscript $i$ stands for the derivative with respect to the in'th spatial
comoving coordinate, and the function $z(\eta)$ is a function of the
cosmological background. If the equation of state of matter is time-independent,
then $z(\eta)$ is proportional to $a(\eta)$. 

For scalar field matter $\varphi(t, x)$ the canonical variable is given by
\begin{equation} \label{Mukhvar}
v \, = \, a \bigl[ \delta \varphi + {{\varphi_0^{'}} \over {\cal H}} \Phi
\bigr] \, ,
\end{equation}
where $\delta \varphi$ is the fluctuation of the matter field,
${\cal H} = a' / a$, and
\begin{equation} \label{zvar}
z \, = \, {{a \varphi_0^{'}} \over {\cal H}} \, .
\end{equation}

The canonical variable $v$ is simply related to the variable ${\cal R}$ which
describes the curvature fluctuations in the comoving coordinate system 
\cite{Bardeen, BST, BK, Lyth} and
which we are interested in computing at late times:
\begin{equation} \label{Rvar}
v \, = \, z {\cal R} \, .
\end{equation}

The equation of motion which follows from the action (\ref{pertact}) is
(in momentum space)
\begin{equation} \label{pertEOM2}
v_k^{''} + k^2 v_k - {{z^{''}} \over z} v_k \, = \, 0 \, ,
\end{equation}
where $v_k$ is the k'th Fourier mode of $v$. We see
that each Fourier mode satisfies a harmonic oscillator
equation of motion with a time-dependent mass,
the time dependence being given by the background
cosmology. The mass term in the above equation is in general
given by the Hubble expansion rate. Thus, we see that
the Hubble radius plays a key role in the evolution of
fluctuations. The mode $k$ whose wavelength at time $t$
is equal to the Hubble radius will be denoted by $k_H(t)$.
it follows that on small length scales, i.e. for
$k > k_H$, the solutions for $v_k$ are constant amplitude oscillations . 
These oscillations freeze out at Hubble radius crossing,
i.e. when $k = k_H$. On longer scales ($k \ll k_H$), there is
a mode of  $v_k$ which scales as $z$. This mode is the dominant
one in an expanding universe, but not in a contracting one.

Canonical quantization of the action for cosmological perturbations
corresponds to imposing canonical commutation relations for
each Fourier mode $v_k$. If we impose vacuum initial conditions
at some time $\eta_i$, this implies
\begin{eqnarray} \label{incond}
v_k(\eta_i) \, = \, {1 \over {\sqrt{2 k}}} \\
v_k^{'}(\eta_i) \, = \, {{\sqrt{k}} \over {\sqrt{2}}} \, \, \nonumber
\end{eqnarray} 

Before applying the above formalism to initial vacuum perturbations
in the matter bounce scenario, we will review how a
scale-invariant spectrum emerges in inflationary cosmology.
The definition of scale-invariance of the curvature power
spectrum is
\be
P_{\cal R} \, \equiv \frac{1}{12 \pi} k^3 |v_k|^2 \, \sim \, k^{n_s - 1} \, \sim \, {\rm const} \, ,
\ee
i.e. $n_s = 1$, where $n_s$ is called the spectral index of scalar metric
fluctuations. Note that the vacuum spectrum, i.e. the spectrum obtained with
the values (\ref{incond}) is not scale invariant. Rather, it is blue with $n_s = 3$
(more power on short wavelengths).

In inflationary cosmology the Hubble radius is constant while the wavelength
of comoving modes expands exponentially. Thus, provided the period of
inflation is sufficiently long, all modes which are currently observed originate
on sub-Hubble scales during inflation. A mode with wavenumber $k$ exits
the Hubble radius at a time $\eta_H(k)$ given by
\begin{equation} \label{Hubble3}
a^{-1}(\eta_H(k)) k \, = \, H \, .
\end{equation}

In inflationary cosmology, any classical fluctuations existing at the beginning
of the period of inflation are red-shifted and (in the absence of a
trans-Planckian mechanism which might re-populate the ultraviolet
modes - see \cite{RHBrev0, Jerome} for discussions of this point) leave
behind a vacuum state. Thus, it makes sense to start fluctuations on
sub-Hubble scales in their vacuum. The fluctuations will oscillate
with constant amplitude while on sub-Hubble scales. However, on
super-Hubble scales $v_k$ will increase as $a(\eta)$. Since
long wavelengths spend more time outside the Hubble radius
they experience a bigger growth. Thus, the final spectrum will be
less blue. When measured at a fixed final time $\eta$, the
increase in the amplitude of $v_k$ will be proportional to
$a(\eta_H(k))^{-1}$, which from (\ref{Hubble3}) is proportional
to $k^{-1}$. Hence, the slope of the power spectrum changes 
by $\delta n_s = -2$, converting the vacuum spectrum into
a scale-invariant one.

We will now see that a similar mechanism converts a vacuum
spectrum into a scale-invariant one in the Matter Bounce
scenario.

\subsection{Vacuum Fluctuations in the Contracting Phase}

In this section we will be following modes which originate as
quantum vacuum fluctuations on sub-Hubble scales at early
times and which cross the Hubble radius during the phase
of matter dominated contraction. The vacuum spectrum
has index $n_s = 3$. To convert it into a scale-invariant
spectrum we require a mechanism which boosts long
wavelengths compared to short wavelengths. Since
$v_k$ grows on super-Hubble scales, such a mechanism
naturally arises in a contracting universe. As we see
below, in a matter-dominated phase of contraction the
boost factor is exactly the right one to turn the vacuum
spectrum into a scale-invariant one \cite{Wands, Fabio1}.

From the equation of motion (\ref{pertEOM2}) it follows that
$v_k$ will oscillate with constant amplitude until the scale
exits the Hubble radius. While the equation of state is
constant in time, we have $z(\eta) \sim a(\eta)$. In a matter-dominated
phase we have $a(\eta) \sim \eta^2$ since $\eta(t) \sim t^{1/3}$.
Hence, it follows by solving (\ref{pertEOM2}) in the limit $k = 0$
that the dominant mode of $v_k$ scales as $\eta^{-1}$. Thus,
the power spectrum of curvature fluctuations at some late time
$\eta$ when the modes of interest to us are outside the
Hubble radius becomes
\bea
P_{{\cal R}}(k, \eta) \, &\sim& k^3 |v_k(\eta)|^2 a^{-2}(\eta) \\
&\sim& \, k^3 |v_k(\eta_H(k))|^2 \bigl( \frac{\eta_H(k)}{\eta} \bigr)^2 \, 
 \nonumber \\
&\sim&  \, k^{3 - 1 - 2} \, \sim \,  {\rm const}  \, , \nonumber 
\eea
where in the first line we have used the relation between ${\cal R}$ and $v$,
in the second the growth rate of $v_k$ on super-Hubble scales,
and in the third line we have made use of the vacuum amplitude of $v_k$ at
the Hubble crossing time $\eta_H(k)$ and the Hubble radius
crossing condition 
\be
\eta_H(k) \, \sim \, k^{-1} \, .
\ee
Thus, we have shown that vacuum perturbations which exit the
Hubble radius during the matter-dominated phase of contraction
acquire a scale-invariant spectrum. After the transition to a
radiation-dominated phase of contraction (if such a phase exists)
all modes are boosted by the same factor, and hence the
scale-invariance of the spectrum persists at least until
the bounce phase begins.

\subsection{Matching Conditions}

So far we have shown that before the bounce the spectrum of
fluctuations of the curvature variable $v$ is scale-invariant.
However, we need to determine the curvature fluctuations in
the expanding phase, i.e. after the bounce. Since the
curvature fluctuations (in the case of adiabatic perturbations)
are constant in time on super-Hubble scales in an expanding
universe, it is sufficient to compute the spectrum immediately
after the bounce at the onset of the period of Standard Cosmology
evolution.

The transfer of curvature fluctuations through a non-singular bouncing
phase is a non-trivial topic. In models with only
adiabatic fluctuations, the bottom line of many detailed studies
is that on length scales which are longer than the time duration of
the bounce the spectral index of the power spectrum of $v$ is
not changed \cite{Cai1, Cai2, Saremi, HLbounce, Stephon, Cai3}
(however, as stressed in \cite{Peter4}, this is not a completely
general result). On
the other hand, in the presence of entropy fluctuations changes
are to be expected \cite{Durrer}. The reason why the analysis
is non-trivial is related to the fact that new physics is required in
order to obtain a non-singular bounce, and when evolving the
fluctuations through the bounce phase the modes associated
with the new physics must be taken into account.

The are two ways of following fluctuations through the bouncing
phase. The first is my explicit numerical integration. In any
given realization of the matter bounce we know what the equations
for the fluctuations are, and we can integrate them. However,
to obtain a good understanding of the results, it is important
to have an analytical method. This method uses matching
conditions at the transition surface from one phase to the
next. Matching conditions were introduced by Israel \cite{Israel}
in the context of matching two solutions of Einstein's equations
across a time-like surface. They were then generalized
to the problem of matching across a space-like boundary 
hypersurface in cosmology by Hwang and Vishniac \cite{HV}
and by Deruelle and Mukhanov \cite{DM}.

The matching conditions state that at the boundary between one
phase and another both the induced metric on the boundary
hypersurface and the extrinsic curvature must be continuous.
Applied to the case of cosmological perturbations, one must
first make sure that the background satisfies the matching conditions,
a point emphasized in \cite{Durrer} (see also the
caveats mentioned in \cite{Peter3}). Once the matching of
the background is successfully achieved, the matching conditions
imply that the metric fluctuation variable $\Phi$ and the 
extrinsic curvature fluctuation must be continuous across the
bounce. If the matching surface is a constant energy density
hypersurface, then the continuity of the extrinsic curvature
implies the continuity of $v$ \cite{Durrer}. 

To model a non-singular bounce, we can merge three
phases, first the contracting phase obeying the equations
of General Relativity, next the bounce phase during which we
can use the parametrization
\be
H(\eta) \, = \, \alpha \eta \, ,
\ee
where $\alpha$ is some constant. We have normalized the
time axis such that the bounce point corresponds to $\eta = 0$.
We must now introduce two matching surfaces \cite{Cai1, Cai2},
first the transition surface between the standard contraction phase
and the bounce phase, and second between the bounce phase
and the standard expanding phase. At the first transition surface
the background is contracting on both sides, and hence the
background matching conditions can easily be satisfied. Similarly,
at the second transition surface the background is expanding on
both sides, and hence the background matching conditions are
again satisfied. 

If both transition surfaces are constant energy density surfaces
(which will automatically be the case if the fluctuations are
adiabatic), then $v$ is continuous across the surface. On
length scales larger than the duration of the bounce phase the
non-trivial evolution during the bounce phase is independent
of $k$ (since the $k^2$ term in the equation of motion is
negligible compared to the term which generalizes the
$z^{\prime \prime}/z$ term in (\ref{pertEOM2}))
and hence, although it can change the amplitude of  the spectrum, 
it will not change the spectral index $n_s$. We
hence conclude that the scale-invariant pre-bounce spectrum
survives the bounce. In Figure 2 we present an example
of the numerically obtained evolution of cosmological
perturbations across a non-singular bounce, in the context
of the model of \cite{Cai3} which will be discussed below.

\begin{figure}
\includegraphics[scale=0.3]{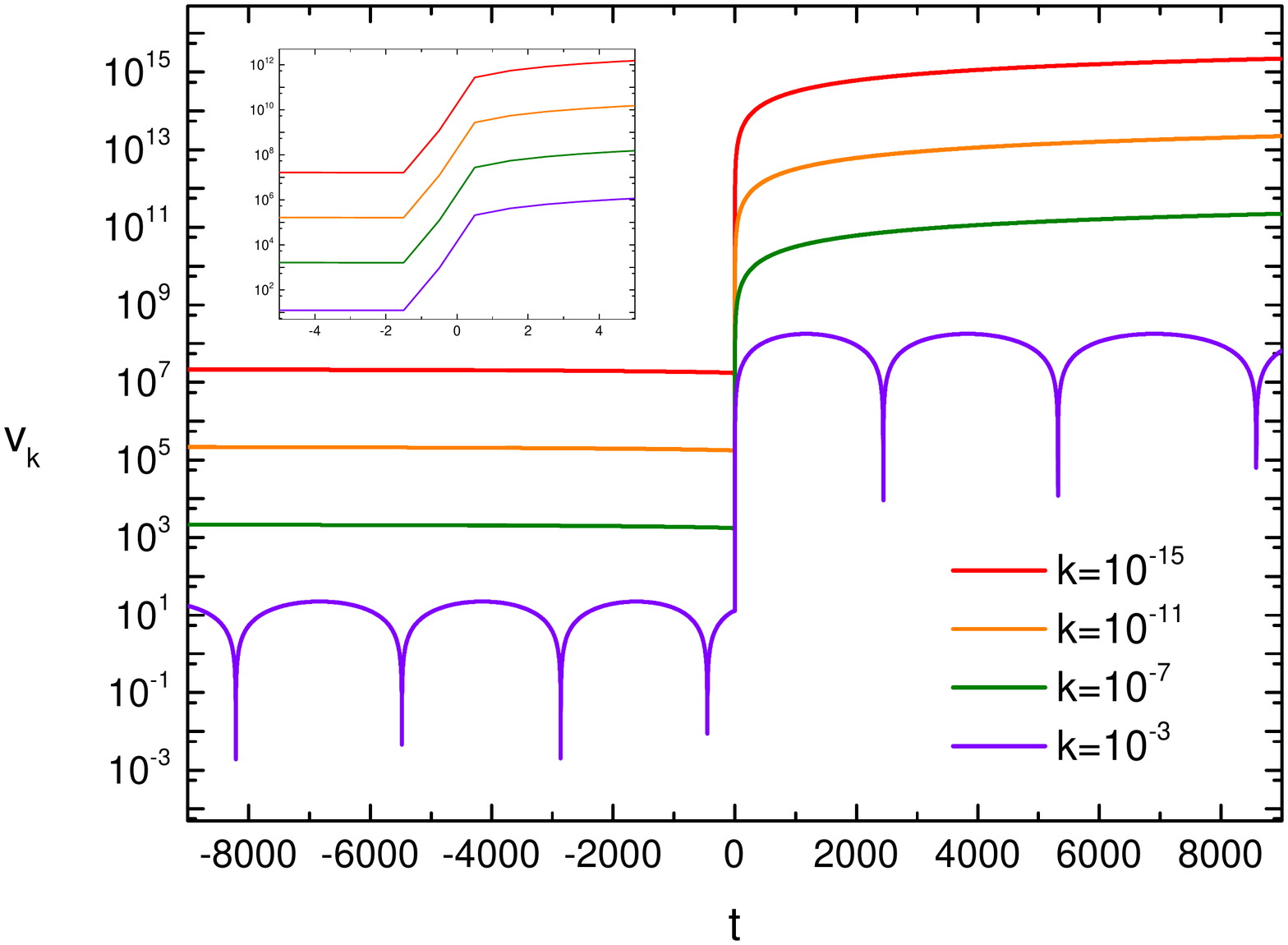}
\caption{Numerical plot of four groups of canonical perturbation modes 
$v_k$ (vertical axis) as functions of cosmic time (horizontal axis). These 
modes are distinguished by the comoving wave number $k$, which are 
$k=10^{-15}$ (in red), $k=10^{-11}$ (in orange), $k=10^{-7}$ (in green), and 
$k=10^{-3}$ (in violet), respectively. The inner insert shows the detailed 
evolution of $v_k$ during the bounce phase. The initial conditions of the 
background field and background parameters are the same as those 
chosen in Fig. 3. All numerical values are in Planck units $M_p$. The 
initial conditions for perturbation modes were chosen to be vacuum initial 
conditions. The amplitude of $v_k$ grows during the bounce phase,
but by the same factor for all long wavelength modes. }
\label{Fig:vsk}
\end{figure}

There is also the possibility of realizing the Matter Bounce
scenario in the context of a singular bounce. One example
is obtained in the context of the string theory backgrounds
with gravitomagnetic fluxes studied in \cite{KPT}. In this
case, there are cosmological solutions which correspond
to a contracting radiation phase glued to an expanding
radiation phase via an S-brane configuration, a configuration
which has vanishing energy density and negative pressure
and thus allows for the violation of the Null Energy Condition
which in turn allows for a cosmological bounce. This model
will be discussed further below. In this context, great care
must be taken when matching the cosmological fluctuations
across the bounce since the extrinsic curvature of the background
changes sign across the S-brane. This issue has been
discussed in \cite{BKPPT} based on the general analysis
of \cite{Durrer}.

Other examples of singular bouncing cosmologies are the
Pre-Big-Bang scenario of \cite{Ven} and the Ekpyrotic 
scenario \cite{Ekp}. As stressed in \cite{Durrer}, in these
cases the application of the matching conditions is
only well defined after careful specification of the matching
surface. This resolved the disagreement between various
results obtained in the literature for the final spectrum
of curvature fluctuations in the Ekpyrotic scenario,
some of which obtained a scale-invariant spectrum
\cite{KOST, TT, Peter2}, and others which did not
\cite{Lyth2, Fabio2, Hwang, Peter}. Calculations done
in the higher-dimensional framework of the Ekpyrotic
scenario show that modes which from the four space-time
dimensional point of view are entropy modes can
mediate a successful transition of the spectrum across
the Ekpyrotic bounce \cite{Thorsten2} (see also
\cite{Riotto, Fabio3, NewEkp}).

After these general remarks on matching conditions
for cosmological fluctuations we will return the
the main development, and we turn to a discussion
of implementations of the Matter Bounce scenario.

\section{Realizing a Matter Bounce}

\subsection{Overview}

As is well known, in order to obtain a bouncing cosmology
it is necessary to either abandon General Relativity as
the theory which describes space and time, or else we
must introduce matter which violates the Null Energy
Condition (NEC). Both options have been used to obtain
realizations of a non-singular bouncing cosmology. 

There is good reason to consider gravitational actions
which differ from that of General Relativity at high
energy scales: since General Relativity is not a
renormalizable quantum theory, we know that in
any approach to quantum gravity there will be terms
in the action which contain higher derivatives. In
this context the singularity theorems of General
Relativity are no longer applicable and it is possible
that nonsingular cosmological solutions emerge.
On the other hand, since higher derivative terms in
the action lead to a larger number of modes, the danger
arises that the singularities will in fact be worse.

There are existence proofs of higher derivative actions
which have non-singular cosmological solutions.
One example is the ``non-singular universe" construction
of \cite{ABM} in which the action is constructed in order
that space-time approached de Sitter space at a certain
limiting curvature. This allows for bouncing cosmological
solutions. Another example is the (non-local) action of
\cite{Biswas} which was constructed to be ghost-free
about Minkowski space-time. Once again, this action
leads to bouncing cosmological solutions. Below,
I will discuss a recent example of a bouncing cosmology
arising from a modified gravitational action, namely
the Ho\v{r}ava-Lifshitz bounce \cite{RHBHL}.

More numerous are examples of bouncing cosmologies
which arise from modifying the matter sector. Below,
I will discuss a few recent examples.

\subsection{Ho\v{r}ava-Lifshitz Bounce}

Ho\v{r}ava-Lifshitz (HL) gravity \cite{HL} is a proposal for a
power-counting renormalizable quantum theory of
gravity in four space-time dimensions. The approach
is conservative in that it uses only the usual metric
variables, but it is radical in the sense that it discards
space-time diffeomorphism and local Lorentz invariance
as underlying symmetries of the theory. HL gravity
postulates the existence of a preferred time direction.
It maintains spatial diffeomorphism invariance, but
space-dependent time reparametrizations are no longer
a symmetry of the theory. Instead, HL gravity postulates
a scaling symmetry which treats space and time anisotropically.
If $x \rightarrow b x$, then there is a scaling symmetry if
$t \rightarrow b^3 t$.

The gravitational action in HL theory is obtained by adding
to the Einstein action all terms which are consistent with the
residual symmetries and which are power-counting renormalizable
with respect to the above-mentioned scaling symmetry. 
This leads to the possibility of adding higher space-derivative
terms to the gravitational action - up to four extra space
derivative terms are allowed. These higher space-derivative
terms can be moved to the right-hand (matter) side of
the Einstein equations. As was realized in \cite{RHBHL},
in the presence of spatial curvature these terms act
as matter with negative effective energy density, thus
violating the NEC and allowing for a bouncing cosmology.

Specifically, if we insert the ansatz for a spatially homogeneous
and isotropic metric into the Ho\v{r}ava-Lifshitz action, we obtain
the following generalization of the Friedmann-Robertson-Walker (FRW)
equations:
\bea
H^2 \, &=& \, - \frac{{\bar k}}{a^2} - \frac{\Lambda_E}{3} - \frac{2 {\bar k}^2 (\zeta + 3 \eta)}{\alpha a^4}
+ \frac{\rho}{6 \alpha} \, , \label{Heq} \\
\bigl[ {\dot H} + \frac{3}{2} H^2 \bigr] \, &=& \, - \frac{{\bar k}}{2 a^2} - \frac{\Lambda_E}{2} + 
\frac{{\bar k}^2 (\zeta + 3 \eta)}{4 \alpha a^4} + \frac{p}{4 \alpha} \, , \label{acceq}
\eea
In the above, $H$ is the usual Hubble expansion rate, $t$ is physical time, an overdot 
indicates the time derivative with respect to $t$, ${\bar k}$ is the spatial curvature constant,
$\Lambda_E$ is the cosmological constant, $\rho$ and $p$ are energy and pressure
densities of matter, respectively, $\alpha = 2 / \kappa^2$ is the coefficient of the standard
kinetic term in the gravitational action, and $\zeta$ and $\eta$ are coefficients of the
higher spatial derivative terms in the potential part of the gravitational action.  

Considering the first FRW equation, we recognize all the terms except the third term
on the right hand side. This term is due to the higher space derivative terms in the
gravitational action. We see that it scales in time like radiation. If the
kinetic term in the gravitational action is close to the form it has in General Relativity,
then $\zeta + 3 \eta > 0$ and thus the new term acts like ghost radiation. Note that
this term is absent if the spatial sections are flat, but that the coefficient does not
depend on whether the spatial curvature is positive or negative.

The bouncing cosmology now arises as follows. We begin with a matter-dominated
phase of contraction. At low curvatures the ghost radiation term is negligible. However,
as $a$ decreases, the importance of ghost radiation relative to matter increases.
Eventually there will be a time when the magnitude of the ghost radiation term
catches up with that of regular matter (for small values of $a$ the first two terms
on the right hand side of the FRW equations are negligible). At this time we
have $H = 0$. From the second FRW equation it follows that ${\dot H} > 0$
at this time. Thus, a cosmological bounce occurs.

\subsection{Quintom Bounce}

Introducing quintom matter \cite{quintom} yields a way of obtaining
a non-singular bouncing cosmology, as discussed in
\cite{Cai1}. Quintom matter is a set of two matter
fields, one of them regular matter (obeying the weak energy
condition), the second a field with opposite sign kinetic
term, a field which violates the energy conditions. We can 
\cite{Cai1} model both matter components with
scalar fields. Let us denote the mass of the regular one 
($\varphi$) by $m$, 
and by $M$ that of the field ${\tilde{\varphi}}$ with 
wrong sign kinetic term. We assume that early in the contracting
phase both fields are oscillating, but that the amplitude $\cal{A}$
of $\varphi$ greatly exceeds the corresponding amplitude
${\tilde{\cal A}}$ of ${\tilde{\varphi}}$ such that the energy density
is dominated by $\varphi$. Both fields will initially be oscillating 
during the contracting phase, and both amplitudes grow at the
same rate. At some point, $\cal{A}$ will become so large 
that the oscillations of $\varphi$ freeze out \footnote{This
corresponds to the time reverse of
entering a region of large-field inflation.}.
Then, $\cal{A}$ will grow only slowly, whereas ${\tilde{\cal A}}$
will continue to increase. Thus, the (negative) energy density
in ${\tilde{\varphi}}$ will grow in absolute values relative to that
of $\varphi$. The total energy density will decrease towards $0$.
At that point, $H = 0$ by the Friedmann equations. 
\be \label{Heq2}
H^2 \, = \, \frac{8 \pi G}{3} \bigl[ \frac{1}{2} {\dot \varphi}^2 - \frac{1}{2} {\dot {\tilde{\varphi}}}^2
+ \frac{1}{2} m^2 \varphi^2 - \frac{1}{2} M^2 {\tilde{\varphi}}^2 + V(\varphi, {\tilde{\varphi}}) \bigr] \, ,
\ee
where $V$ is the potential of the two fields.
From the second FRW equation 
\be \label{dotHeq}
{\dot H} \, = \,  - 4 \pi G \bigl({\dot \varphi}^2 - {\dot {\tilde{\varphi}}}^2 \bigr) 
\ee
it follows that since ${\tilde{\varphi}}$ is oscillating while $\varphi$ is
only slowly rolling that ${\dot{H}} > 0$ when $H = 0$. Hence,
a non-singular bounce occurs. The Higgs sector of the
Lee-Wick model \cite{LW} provides a concrete realization of 
the quintom bounce model, as studied in \cite{Cai2}.
Quintom models like all other models with negative sign kinetic
terms suffer from an instability problem \cite{ghost} in the
matter sector and are hence problematic. In addition, they are
unstable against the addition of radiation (see e.g. \cite{Karouby})
and anisotropic stress.

\subsection{Ghost Condensate Bounce}

An improved way of obtaining a non-singular bouncing cosmology
using modified matter is by using a ghost condensate field \cite{Chunshan}
(see also \cite{GBounce} for related work).
The ghost condensation mechanism is the analog of the Higgs
mechanism in the kinetic sector of the theory. In the Higgs
mechanism we take a field $\varphi$ whose mass when evaluated
at $\varphi = 0$ is tachyonic, add higher powers of $\varphi^2$ to the
potential term in the Lagrangian such that there is a stable
fixed point $\varphi = v \neq 0$, and thus when expanded about
$\varphi = v$ the mass term has the ``safe" non-tachyonic sign.
In the ghost condensate construction we take a field $\varphi$
whose kinetic term
\be
X \, \equiv \,  - g^{\mu \nu} \partial_{\mu} \phi \partial_{\nu} \varphi 
\ee
appears with the wrong sign in the Lagrangian. Then, we add
higher powers of $X$ to the kinetic Lagrangian such that there
is a stable fixed point $X = c^2$ and such that when expanded
about $X = c^2$ the fluctuations have the regular sign of the kinetic
term:
\be
{\cal L} \, = \, \frac{1}{8} M^4 \bigl( X - c^2 \bigr)^2 - V(\varphi) \, ,
\ee
where $V(\varphi)$ is a usual potential function, $M$ is a 
characteristic mass scale and the dimensions of $\varphi$
are chosen such that $X$ is dimensionless.

In the context of cosmology, the ghost condensate is
\be
\varphi \, = \, c t 
\ee
and breaks local Lorentz invariance. Now let us expand the
homogeneous component of $\varphi$ about the ghost condensate:
\be
\varphi(t) \, = \, c t + \pi(t) \, .
\ee
If ${\dot \pi} < 0$ then the gravitational energy density is negative,
and a non-singular bounce is possible. Thus, in \cite{Chunshan}
we constructed a model in which the ghost condensate field
starts at negative values and the potential $V(\varphi)$ is
negligible. As $\varphi$ approaches $\varphi = 0$ it encounters a
positive potential which slows it down, leading to ${\dot \pi} < 0$
and hence to negative gravitational energy density. Thus,
a non-singular bounce can occur. We take the potential
to be of the form
\be
V(\varphi) \, \sim \, \varphi^{- \alpha}
\ee
 for $|\phi| \gg M$, where $M$ is the mass scale above which
 the higher derivative kinetic terms are important. For
 sufficiently large values of $\alpha$, namely
 \be
 \alpha \, \geq \, 6 \, ,
 \ee
 the energy density in the ghost condensate increases faster
 than that of radiation and anisotropic stress at the universe
 contracts . Hence, this bouncing cosmology is stable
 against the addition of radiation and anisotropic stress.
 
 However, the model is still unstable to the presence of 
 anisotropic stress during the long periods of matter and
 radiation domination which precede the bounce phase.
 We now turn to a model which also solves this
 problem. 

\subsection{Ekpyrotic Bounce}

One of the key advantages of the Ekpyrotic bouncing model \cite{Ekp}
is that anisotropies decay relative to the dominant component of
the energy density during the period of contraction \cite{noBKL}.
The Ekpyrotic scenario is obtained by introducing a scalar field
with negative exponential potential. While the scalar field is
rolling down the potential towards $\varphi = 0$, the equation
of state parameter $w \, \equiv \, p / \rho$ satisfied
\be
w \, \gg \, 1 \, ,
\ee
since there is a partial cancellation between the positive kinetic
energy density and the negative potential. From the continuity
equation it hence follows that the energy density $\rho(\varphi)$
increases faster than that in anisotropies.

The spectrum of adiabatic fluctuations in the Ekpyrotic scenario
is blue \cite{Lyth2, Fabio2}, and hence additional inputs (e.g.
entropy fluctuations) are required to obtain a scale-invariant
spectrum of curvature fluctuations.

The Ekpyrotic Bounce model of \cite{Cai3} combines the
advantages of the ``Matter Bounce" scenario in terms
of obtaining a scale-invariant spectrum of adiabatic
fluctuations with the advantages of the Ekpyrotic
model in solving the anisotropy problem of a contracting universe.
The model is obtained by introducing a scalar field $\varphi$
which combines the ghost condensate or
Galileon mechanisms for obtaining a non-singular bounce
discussed in the previous subsection with a negative exponential
potential which renders $\rho(\varphi)$ to be dominant over
anisotropies at later stages during the contracting phase.

The model of \cite{Cai3} is based on
the most general form of single scalar field Lagrangian giving rise 
to second-order 
field equations in  four-dimensional spacetime
\cite{Horndeski:1974, Deffayet:2011gz}
\begin{eqnarray}\label{Lagrangian}
 {\cal L} \, = \, K(\varphi, X) + G(\varphi, X) g^{\mu \nu} \partial_{\mu} \partial_{\nu} \varphi +  ... \, ,
\end{eqnarray}
where $K$ and $G$ are functions of a dimensionless scalar field $\varphi$ and its 
canonical kinetic term $X$, 
and where we have not written down a number of higher order operators which
are not important at lower energy scales (we shall construct our bounce to
be consistent with the condition that these terms remain negligible).

Following the idea behind the ghost condensate bounce of
\cite{Chunshan} we assume that the term $K$ is given by
\begin{eqnarray}\label{Kessence}
 K(\varphi, X) = M_p^2 [1-g(\varphi)]X + \beta X^2 - V(\varphi)~,
\end{eqnarray}
where we introduce a positive-definite parameter $\beta$ so that the kinetic term is bounded 
from below at high energy scales. For $g > 1$ a ghost condensate ground state with
$X \neq 0$ can arise. This is the first key principle of our model.
Note that the first term of $K$ involves $M_p^2$ since in 
the present paper we adopt the convention that the scalar field $\varphi$ is dimensionless. 

The second key principle of our model is to introduce a non-trivial potential 
$V$ for $\varphi$ which  is chosen such that Ekpyrotic contraction is possible. 
In the specific model which we will discuss 
in the following, the scalar field evolves monotonically from a negative large values to a 
positive large value. The function $g(\phi)$ is chosen such that a phase of ghost 
condensation only occurs during a short time when $\phi$ approaches $\phi=0$. This 
requires the dimensionless function $g$ to be smaller than unity when $|\phi| \gg 1$ 
but larger than unity when $\phi$ approaches the origin.  

We choose $G$ to be a simple function of only $X$:
\begin{eqnarray}\label{Galileon}
 G(X) = \gamma X~,
\end{eqnarray}
where $\gamma$ is a positive number. This term is introduced
to stabilize gradients.

The potential is chosen such that during the approach to $\varphi = 0$
from negative values we have Ekpyrotic contraction:
\begin{eqnarray}
 V(\phi) \, = \, -\frac{2V_0}{e^{-\sqrt{\frac{2}{q}}\varphi}+e^{b_V\sqrt{\frac{2}{q}}}\varphi}~,
\end{eqnarray}
where $V_0$ is a constant.

Based on the previous discussion, the function $g(\varphi)$ is chosen in the
following form
\begin{eqnarray}
 g(\phi) = \frac{2g_0}{e^{-\sqrt{\frac{2}{p}}\phi}+e^{b_g\sqrt{\frac{2}{p}}\phi}}~,
\end{eqnarray}
where $g_0$ is a positive constant greater than $1$.

The background equations of motion which follow from the Lagrangian (\ref{Lagrangian})
with the choices of the various free functions discussed above can
be solved analytically using approximate methods. They can also be solved
numerically. The results of the numerical integration are shown in Figure 3.
\begin{figure}
\includegraphics[scale=0.3]{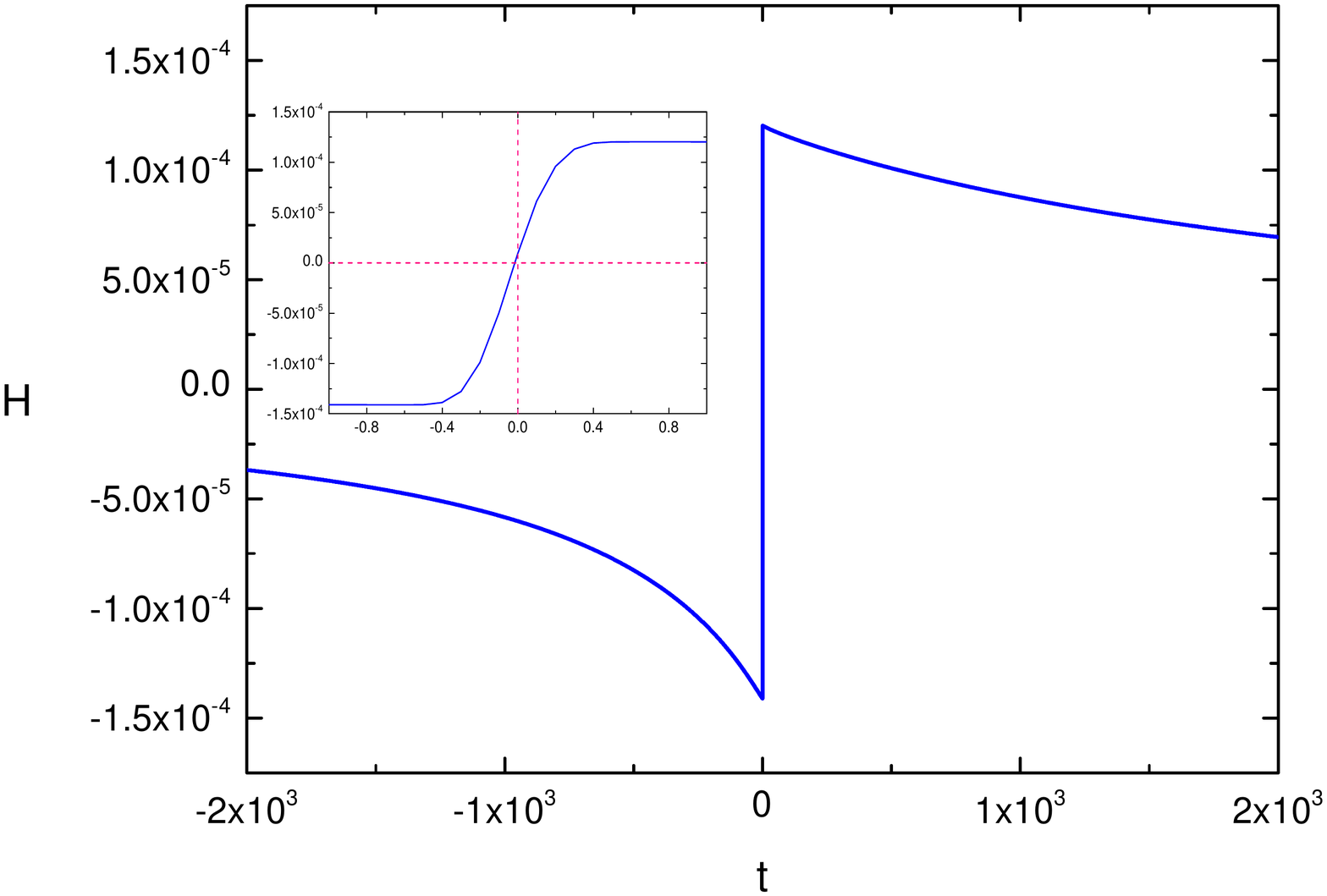}
\caption{Numerical plot of the Hubble parameter $H$ (vertical axis) as a function 
of cosmic time (horizontal axis). The main plot shows that a non-singular bounce occurs,
and that the time scale of the bounce is short. The inner 
insert shows a blowup of the Hubble parameter $H$ during the bounce phase. 
All numerical values are in Planck units $M_p$.  The parameters were chosen
to be $V_0 = 10^{-7}~,~~g_0 = 1.1~,~~\beta = 5~,~~\gamma = 10^{-3}~,
 b_V = 5~,~~b_g = 0.5~,~~p = 0.01~,~~q=0.1$. }
\label{Fig:Hubble}
\end{figure}
As is evident, a successful non-singular bounce occurs.

We have also studied \cite{Cai3} the evolution of cosmological fluctuations
through the bounce phase and verified that on large scales the spectral
index does not change (see Figure 2).

\subsection{Stringy S-Brane Bounce}

All the models with modified matter we have considered so
far are toy models in that they do not arise from a theory
which is well behaved in the ultraviolet. We conclude
this section by mentioning a model which originates from
a theory which is well behaved in the ultraviolet, namely
superstring theory. 

It has long been realized that string
theory has the potential of solving the singularity problem
of Standard Big Bang cosmology. As is well known
\cite{Hagedorn}, there is a maximal temperature of a
gas of weakly coupled strings in thermal equilibrium,
the Hagedorn temperature $T_H$. In the context
of a toroidal compactification of Heterotic string theory,
it was realized \cite{SGC} that the temperature remains
smaller than $T_H$ throughout. However, there are
potential instabilities of string theory near the
Hagedorn temperature.

Recently \cite{KPT}, ingredients to resolve both the 
Hagedorn instability as well as the initial curvature singularity of
string cosmology were proposed.  These ingredients are realized in a class of 
${\cal{N}}=(4,0)$ Type II superstring models where the presence of non-trivial 
``gravito-magnetic" fluxes modifies the thermal vacuum by adding non-trivial momentum
and winding charges, lifting the Hagedorn instabilities.
If $R_0$ denotes the radius of the Euclidean time circle, 
the string partition function $Z$ has a thermal duality symmetry 
about a critical value $R_c$, which is of the order of the string scale:  
\be
\label{dual}
Z(R_0) \, = \, Z(R_c^2 / R_0) \, .
\ee
The inverse temperature $\beta(R_0)$ measured in the string frame is then 
proportional to $R_0$ or its dual value in the asymptotic regimes:
\be
\label{beta(R)}
\beta \sim 2\pi R_0\;\;\mbox{  as  }\; \;R_0\gg R_c\, , \quad 
\beta \sim 2\pi\,  {R_c^2\over R_0}\;\;\mbox{  as  }\; \;R_0\ll R_c\, .
\ee
Thus the temperature acquires a maximal critical value.

In the adiabatic approximation, cosmological
dynamics very different from the usual particle physics induced
dynamics can be obtained. One possibility is a solution in which 
$R_0$ runs from $0$ to $\infty$. At sub-critical values of $R_0$,
the Universe can be taken to be in a contracting phase.  At the critical value of $R_0$,
states characterized by non-trivial winding and momentum charges along the 
Euclidean time circle become massless and give rise to a Euclidean gauge 
theory. At the level of the four-dimensional low energy
effective field theory of dilaton-gravity, these massless states 
lead to extra terms in the effective action which
correspond \cite{KPT} to an S-brane configuration (see
e.g. \cite{sbrane} for early references on S-branes), and  
the effective action is that of dilaton-gravity coupled to a
thermal gas of strings plus the S-brane configuration. An
S-brane acts like a topological defect in space-like direction.
Its energy density vanishes in the same way that the
transverse pressure of a regular defect vanishes. The
pressure is negative. Hence,
the S-brane provides the violation of the null energy 
condition (NEC) which leads to a smooth transition from a contracting
phase to an expanding phase (when cosmological evolution is
viewed in the Einstein frame).  Thus, for $R_0 > R_c$ the Universe 
expands. Thus, the setup of \cite{KPT} provides a realization
of a bouncing cosmology. The metric in this setup is continuous,
but the derivative of the metric diverges. 

At physical temperatures much lower than the critical value,
the partition function of the string gas has the equation of
state of radiation as long as supersymmetry is unbroken
\footnote{Note that the dilaton approaches a constant at
low temperatures, and thus equations of motion of Einstein
gravity for cosmological perturbations are applicable.}.
However, at low temperatures we expect supersymmetry
to be broken, thus generating masses and leading to the
emergence of a phase of matter domination at low
temperature.

In \cite{BKPPT}, the evolution of cosmological fluctuations
in this bouncing cosmology has been studied in detail.
The background cosmology starts with a matter-dominated
phase of contraction. At a certain temperature $T_{\rm SUSY}$
supersymmetry is restored, and a transition to a phase
of radiation domination will occur. Once the temperature
rises to the dual value, the S-brane will be reached. The
S-brane mediates the transition to an expanding radiation
phase. The temperature will decrease, and below $T_{\rm SUSY}$
supersymmetry will break, allowing for the generation of
a component of pressureless matter which dominates the
late-time dynamics. 

This background thus provides
the requirements to implement the Matter Bounce scenario:
we begin with quantum vacuum perturbations on sub-Hubble
scales in the contracting phase. Scales which exit the
Hubble radius before the onset of the radiation phase
acquire a scale-invariant spectrum. In the absence of
entropy fluctuations the perturbations are matched across
the S-brane with no change in the spectral index.
Hence, at late times in the expanding phase a scale-invariant
spectrum of curvature fluctuations results.

\section{Conclusions}
 
The ``Matter Bounce" is an alternative to cosmological inflation
as a mechanism for generating an almost scale-invariant
spectrum of cosmological fluctuations. As in the case of
inflationary cosmology, the spectrum has a slight red
tilt since smaller scales exit the Hubble radius when the
radiation component of matter is more important and the
vacuum slope of the spectrum (which corresponds to $n_s = 3$)
is rearing its head. In fact, there is a transition to an
$n_s = 3$ spectrum on small scales which exit the
Hubble radius in the radiation phase of contraction.
The current observational constraints on a spectrum with
such a transition from $n_s = 1$ on large scales to
$n_s = 3$ on small scales were studied in \cite{LiHong},
with the result that the power spectrum of quasar absorption
lines already mildly constrains a model with no entropy
production during the bounce. If there is entropy production
during the bounce, then the radiation phase of contraction
is shorter than the radiation phase of expansion and the
observational constraints recede.

The spectrum of gravitational radiation produced in a
matter bounce is also scale-invariant. At first glance, there
will be less enhancement of the amplitude during the
bounce phase than for scalar metric fluctuations,
and hence the predicted tensor to
scalar ratio $r$ will be consistent with current constraints.
However, a detailed study of this issue is needed.

Since the curvature fluctuation ${\cal R}$ increases on
super-Hubble scales during the contracting period, terms
in the general expression for the bispectrum (three
point function of the curvature fluctuation variable) derived in
\cite{Malda} will be important, leading to a shape of
the bispectrum which is different from the one obtained
in simple inflationary models \cite{bispectrum}. 
Since the analog of the inflationary slow-roll parameter is 
of order unity in the case of the Matter Bounce, the predicted
amplitude of the bispectrum will be of order unity \cite{bispectrum}.
Thus, amplitude and shape of the bispectrum provide a way
to differentiate between the Matter Bounce and simple inflationary
models.

Possibly the most serious problem which the Matter Bounce scenario
faces is the anisotropy problem (see e.g. \cite{BingKan}). 
The only solution of this problem which we know of now is to
postulate a phase of Ekpyrotic contraction following the initial
matter contraction phase. In the model of \cite{Cai3} this
solves the anisotropy problem, whereas in the original
New Ekpyrotic scenario the anisotropy problem may
re-appear during the bounce phase \cite{BingKan}.

I have focused on single bounce models. Cyclic bouncing
background cosmologies face several challenges. In
addition to the ``heat death" problem (entropy increasing
from cycle to cycle), there is the problem that the dynamics
of cosmological fluctuations breaks the cyclicity of the model
since fluctuations grow on super-Hubble scales during the
periods of contraction and lead to a jump in the index
$n_s$ of the power spectrum by $\delta n_s = -2$ from
cycle to cycle \cite{processing}. 
Thus, the model loses predictability \footnote{Note
that the cyclic version of the Ekpyotic scenario \cite{Ekpcyclic}
does not face this problem since the evolution of the
scale factor of our universe is not cyclic in the model,
only the separation between the branes in the higher-dimensional
framework of the model.}. 

\begin{acknowledgments}

I wish to thank my collaborators on whose work I have drawn. In particular,
I thank Dr. Yifu Cai for collaboration on many of these topics
and for allowing me to make use of the figures. Special thanks to
Professor Xinmin Zhang for collaboration over a period of many
years and for wonderfully hosting me repeatedly at IHEP in
Beijing. This work was supported in part by an NSERC Discovery
Grant and by funds from the Canada Research Chair program. 

\end{acknowledgments}


\end{document}